\documentclass[12pt]{article}
\usepackage{amsmath,amssymb}
\usepackage[dvips]{graphicx}
\newcommand{\be}{\begin{equation}}
\newcommand{\ee}{\end{equation}}
\newcommand{\bea}{\begin{eqnarray}}
\newcommand{\eea}{\end{eqnarray}}

\def \a {$\alpha$}
\def \aone {$\alpha_1$}
\def \atwo {$\alpha_2$}
\def\Gf {Green's function}
\def\Gfs {Green's functions}
\def \z {$Z_3^{-1}$}

\begin{document}
\renewcommand {\theequation}{\thesection.\arabic{equation}}
\renewcommand {\thefootnote}{\fnsymbol{footnote}}
\vskip1cm
\begin{flushright}
\end{flushright}
\vskip1cm
\begin{center}
{\Large\bf Gauge-Dependence of Green's Functions\\\vskip 0.3 cm in
QCD and QED}

\vskip .7cm

{\bf{\large{K. Nishijima$^a$ and A. Tureanu$^b$}}\\

\vskip .7cm

{\it $^a$Department of Physics, University of Tokyo 7-3-1 Hongo,\\
Bunkyo-ku, Tokyo 113-0033, Japan\\
$^b$High Energy Physics Division, Department of Physical Sciences,
University of Helsinki\\and Helsinki Institute of Physics, P.O. Box
64, FIN-00014 Helsinki, Finland}}

\end{center}

\vskip1cm
\begin{abstract}

When all Green's functions are known in a given gauge we may raise a
question of whether it is possible or not to derive the
corresponding ones in a different gauge. The answer is negative in
QCD but affirmative in QED provided that we confine ourselves to the
covariant gauge characterized by a gauge parameter \a. We shall
discuss the physical significance of this conclusion.

\end{abstract}

\vskip1cm


\section{Introduction}
\setcounter{equation}{0}

In quantizing a gauge-invariant Lagrangian we encounter a well-known
difficulty in finding the canonical conjugate of the time-component
of the gauge field. This is a reflection of the non-uniqueness of
the solution of the gauge field equation due to the gauge freedom.
This difficulty has been resolved in QED by Fermi's introduction of
the gauge-fixing term into the Lagrangian that provides us with the
lacking canonical conjugate.

Once the theory is quantized by this method, however, the
introduction of indefinite metric is indispensable since the vector
field is obliged to inherit it from the Minkowski metric. This means
that the state vector space resulting from the quantization of the
gauge field is larger than needed for physical interpretation, and
we have to pick out its physical subspace by introducing a
subsidiary condition. This condition plays a dual role of
eliminating the indefinite metric inherent in the state vector space
as well as of recovering the classical gauge field equation, not
modified by the gauge-fixing term, in the physical subspace.

In the present paper we shall confine ourselves to the so-called
covariant gauge specified by a parameter \a\ called the gauge
parameter. Quantized gauge theories are no longer invariant under
local gauge transformations because of the presence of the
gauge-fixing term, but it so happens that the resulting theory is
invariant under new global transformations called the BRS
transformations \cite{brs}.

In Section 2 we shall reinstate the essence of the BRS
transformations in connection wit the gauge-dependence of Green's
functions and shall clarify the condition under which \Gfs\ become
gauge-independent. In tackling the problem of gauge-dependence the
renormalization group (RG) approach \cite{G-ML,BS} is useful and it
is briefly recapitulated in Section 3 with an emphasis on the gauge
field propagator.

The question of gauge-dependence was once raised by Shirkov in
connection with the use of RG in perturbative QCD \cite{Shirkov}.
The results depend sensitively on the renormalization scheme so that
we propose here a scheme of making the beta-function independent of
the gauge parameter \a.

In Section 4 the RG equations for the running coupling constant and
gauge parameter are solved and their ultra-violet asymptotic limits
are studied to answer the question of gauge-dependence of Green's
functions. For this purpose we derive a sum rule \cite{KN96,NT}
which enables us to express the renormalization constant \z\ in a
simple form.

\Gfs\ defined as the vacuum expectation values of the time-ordered
products of BRS invariant operators are always independent of \a,
but those made up  of BRS variant ones are \a-dependent. For the
latter an important question is whether it is possible or not to
continue \Gfs\ analytically as functions of \a. When it is possible
to continue them from \aone\ to \atwo, we say that \aone\ and \atwo\
are connected. Then, mutually connected values of \a\ form a set
called an equivalence class of gauges, and we may ask how many
equivalence classes of gauges there are in QCD and also in QED. The
answer is that three in QCD and one in QED, respectively. QED is
simple, but QCD is complicated and we may ask what the multiplicity
of equivalence classes would mean. We shall discuss the physical
significance of these results in connection with the gluon mass.

In Section 5 this quest is further pursued by extending the gauge
parameter \a\ into the complex plane.

\section{BRS Invariance}
\setcounter{equation}{0}

Local gauge transformations in classical gauge theory are replaced
by global BRS transformations and we shall briefly recapitulate
their properties

\vskip1cm {\it BRS transformations}

The standard Lagrangian density of a gauge theory, say QCD, is given
by
\be\label{lagrangian}{\cal L}={\cal L}_{{inv}}+{\cal L}_{{gf}}+{\cal
L}_{{FP}}\ee
where ${\cal L}_{{inv}}$ denotes the classical gauge-invariant part,
${\cal L}_{{gf}}$ the gauge-fixing terms and ${\cal L}_{{FP}}$ the
Faddeev-Popov (FP) ghost term characteristic of non-Abelian gauge
theories:
\bea {\cal L}_{{inv}}&=&-\frac{1}{4}F_{\mu\nu}\cdot
F_{\mu\nu}-\bar\psi(\gamma_\mu D_\mu+m)\psi\,,\cr
{\cal L}_{{gf}}&=&\partial_\mu B\cdot A_\mu+\frac{1}{2}\alpha B\cdot
B\,,\cr
{\cal L}_{{FP}}&=&i\partial_\mu \bar c\cdot D_\mu
c\,.\label{lagr_terms} \eea
in the customary notation. The gauge parameter is denoted by \a\ and
$D_\mu$ represents the covariant derivative whose explicit forms are
given by
\bea D_\mu\ \psi&=&(\partial_\mu-igT\cdot A_\mu)\psi\,,\cr
 D_\mu\ c&=&\partial_\mu c+gA_\mu\times c\,.\label{cov_deriv}\eea

The BRS transformations of the gauge field $A_\mu$ and the quark
field $\psi$ are defined by replacing the infinitesimal gauge
function by the FP ghost field $c$ or $\bar c$ in their respective
infinitesimal gauge transformations:
\bea \delta A_\mu&=&D_\mu c\,,\ \ \ \delta\psi=ig(c\cdot
T)\psi\,,\cr
\bar\delta A_\mu&=&D_\mu \bar c\,,\ \ \ \bar\delta\psi=ig(\bar
c\cdot T)\psi\,.\label{brs_transf} \eea
For the auxiliary fields $B$, $c$ and $\bar c$ we require
\be \delta{\cal L}=\bar\delta{\cal L}=0\,, \ee
then we find
\bea \delta\,B=0\,,\ \ \ \delta\,\bar c=i B\,,\ \ \
\delta\,c=-\frac{1}{2}g \,(c\times c)\,,\cr
\bar\delta\,\bar B=0\,,\ \ \ \bar\delta\,c=i \bar B\,,\ \ \
\bar\delta\,\bar c=-\frac{1}{2}g\,(\bar c\times \bar c)\,, \eea
where $\bar B$ is defined by
\be B+\bar B-ig(c\times \bar c)=0\,.\ee

In general the BRS transforms of a field $\phi$ are given in terms
of the BRS charges $Q_B$ and $\bar Q_B$ by
\bea \delta\,\phi=i[Q_B,\phi]_\mp,\ \ \ \bar\delta\,\phi=i[\bar
Q_B,\phi]_\mp\,,\\\label{brs_phi}
Q_B^2={\bar Q}_B^2=Q_B\bar Q_B+\bar Q_BQ_B=0\,.\label{brs_charge}
\eea
We choose the $-(+)$ sign in (\ref{brs_phi}) when $\phi$ is even
(odd) in the ghost fields $c$ and $\bar c$ that are anticommuting
hermitian scalar fields.

The sum of the gauge-fixing and the FP ghost terms can be expressed
as
\be{\cal L}_{{gf}}+{\cal L}_{{FP}}=\delta(-i\partial_\mu\bar c\cdot
A_\mu-\frac{i}{2}\alpha\,\bar c \cdot B)\ee
and evidently we have
\be \delta{\cal L}_{inv}=0\,. \ee
Namely, ${\cal L}_{inv}$ is closed and ${\cal L}_{{gf}}+{\cal
L}_{{FP}}$ is exact, and
\be \delta{\cal L}=0\,. \ee

\vskip1cm {\it BRS Cohomology} \cite{KN96}

The quantization of the gauge field and the introduction of the
auxiliary fields $B$, $c$ and $\bar c$ introduce indefinite metric
into the state vector space ${\cal V}$.

A physical state $|f\rangle$ is defined by the following subsidiary
condition:
\be Q_B\,|f\rangle=0\,,\ \ \ \ |f\rangle\in{\cal V}\,. \ee
In particular, the vacuum state $|0\rangle$ is physical,
\be Q_B\,|0\rangle=0\,. \ee
The physical subspace ${\cal V}_{phys}$ is then defined by
\be{\cal V}_{phys}=\{|f\rangle|\ \ Q_B|f\rangle=0\,,\ \
|f\rangle\in{\cal V}\}\,.\ee
It is essentially a collection of closed states with respect to the
nilpotent operator $Q_B$. We also introduce a subspace ${\cal V}_d$
defined by
\be {\cal V}_{d}=\{Q_B|f\rangle|\ \ |f\rangle\in{\cal V}\}\,.\ee
This is a collection of exact states with respect to $Q_B$ and the
Hilbert space $\cal H$ is defined as the BRS cohomology by
\be {\cal H}= {\cal V}_{phys}/{\cal V}_{d}\,.\ee
Then let us consider a set of closed operators $A,\ B, \cdots$
satisfying
\be\label{closure}\delta A=\delta B=\cdots =0\,,\ee
then
\be\label{M variation}\langle 0|\delta M\cdot
AB\cdots|0\rangle=\langle 0|\delta (MAB\cdots|0\rangle=0\,,\ee
since the vacuum state is physical.

Let ${\cal L}_I$ and ${\cal L}_{II}$ be two BRS invariant Lagrangian
densities, namely,
\be \delta{\cal L}_I=\delta{\cal L}_{II}=0\,. \ee
Furthermore, let us assume that their difference be exact:
\be\label{exact difference} \Delta {\cal L}=\delta{\cal
L}_{II}-\delta{\cal L}_I=\delta M\,. \ee
For instance, Lagrangian densities corresponding to two distinct
values of \a\ in (\ref{lagr_terms}) satisfy these two conditions
since we find
\be M=-\frac{1}{2}i\,\Delta\alpha\,(\bar c\cdot B)\,.\ee
Then we introduce time-ordered \Gfs\ in two gauges given above.

In the gauge I we have the following path integrals:
\bea \langle AB\cdots\rangle_I=\frac{1}{N_I}\int {\cal D}(path)\,
AB\cdots \exp(iS_I)\\
N_I=\int {\cal D}(path)\,\exp(iS_I)\,. \eea
${\cal D}(path)$ denotes the path integral over all the field
variables, and we have similar expressions in the gauge II, and the
difference between the two actions is given by
\be\Delta S=S_{II}-S_I=\int d^4 x\,\Delta{\cal L}=\int d^4 x\,\delta
M\,.\ee
\Gfs\ in the gauge II can be expressed as
\be\label{greenII}\langle AB\cdots\rangle_{II}=\frac{\langle
AB\cdots\exp(i\,\Delta S)\rangle_I}{\langle\exp(i\,\Delta
S)\rangle_I}\,.\ee
Now we expand the denominator on the r.h.s. of (\ref{greenII}) in
powers of $\Delta S$ and use (\ref{exact difference}) and (\ref{M
variation}) to obtain
\be\langle\exp(i\,\Delta S)\rangle_I=1\,.\ee
Thus (\ref{greenII}) reduces to
\be\label{2.28}\langle AB\cdots\rangle_{II}={\langle
AB\cdots\exp(i\,\Delta S)\rangle_I}\,.\ee
The r.h.s. can be expanded in powers of $\Delta
\alpha=\alpha_{II}-\alpha_I$. When this expansion converges this is
an analytic continuation of \Gfs\ as functions of \a. When this is
the case we may say that the two values of \a, \aone\ and \atwo, are
connected and the set of \a\ mutually connected will be called an
equivalence class of gauges. The question of how many classes there
are in a given gauge theory will be discussed in a later section.

When all the operators are closed, satisfying (\ref{closure}), we
obtain, with the help of (\ref{M variation}) the equality
\be\label{2.29}\langle AB\cdots\rangle_{II}={\langle
AB\cdots\rangle_I}\,.\ee
This shows that \Gfs\ cosntructed in terms of closed operators alone
are gauge-independent. The $S$-matrix elements for observable
hadronic processes are obtained by applying the LSZ reduction
formula \cite{LSZ} to \Gfs\ defined in terms of BRS invariant
operators so that they are independent of the choice of the gauge
parameter \a. Of course, this statement does not apply to \Gfs\
constructed in terms of BRS variant operators, such as the gauge
field propagator which we shall investigate in a later section.

\section{Renormalization Group}
\setcounter{equation}{0}

Once Shirkov emphasized the gauge.dependence of the RG treatment in
perturbative QCD \cite{Shirkov} since the results are sensitive to
the renormalization scheme. He even presented an example in which
asymptotic freedom is valid only for positive values of \a\ but not
for negative ones. This is apparently due to the \a\ dependence of
the beta-function. Therefore, we shall first briefly review a
renormalization scheme which makes the beta-function independent of
the gauge parameter \cite{NT}.

First we shall refer to (\ref{2.28}) which has been derived in the
unrenormalized version. The field operators $A_\lambda$ and $B$ and
the gauge parameter \a\ are multiplicatively renormalized:
\be\label{3.1} A_\lambda^{(0)}=Z_3^{1/2}(\alpha)\,A_\lambda\,,\ \ \
B^{(0)}=Z_3^{-1/2}(\alpha)\,B\,,\ \ \
\alpha^{(0)}=Z_3(\alpha)\,\alpha\,,\ee
where the superscript $(0)$ is attached to the unrenormalized
expressions.

As a special case of (\ref{2.28}) we shall choose
$\alpha_{II}=\alpha$ and $\alpha_I=0$, then we have
\be\label{3.2}\langle AB\cdots\rangle_{\alpha}=\left\langle
AB\cdots\exp\left(\frac{i\alpha}{2}\int d^4 x\,B(x)\cdot
B(x)\right)\right\rangle_0\,,\ee
in the unrenormalized version. The renormalized version of
(\ref{3.2}) takes exactly the same form because of the identity
\be\alpha^{(0)}B^{(0)}\cdot B^{(0)}=\alpha\,B\cdot B\,.\ee
Then we introduce RG equations in the Landau gauge, $\alpha=0$. For
the renormalized version of \Gfs\ we have an equation of the
following form:
\be ({\cal D}_0+\gamma_G)\langle AB\cdots\rangle_0=0\,, \ee
where $\gamma_G$ is the anomalous dimension of \Gf\ $\langle
AB\cdots\rangle_0$ in the Landau gauge and
\be {\cal
D}_0=\mu\frac{\partial}{\partial\mu}+\beta(g)\frac{\partial}{\partial
g}\,. \ee
The anomalous dimension of the gauge field in the Landau gauge,
which is obviously \a-independent, is denoted by $\gamma_V$.

Then the l.h.s. of (\ref{3.2}), renormalized in the Landau gauge,
satisfies
\be ({\cal D}_0+\gamma_G)\langle
AB\cdots\rangle_\alpha=2\gamma_V\left\langle
AB\cdots\exp\left(\frac{i\alpha}{2}\int d^4 x\,B(x)\cdot
B(x)\right)\right\rangle_\alpha\,. \ee
On the other hand, we also have
\be \alpha\frac{\partial}{\partial\alpha}\langle
AB\cdots\rangle_\alpha=\left\langle
AB\cdots\exp\left(\frac{i\alpha}{2}\int d^4 x\,B(x)\cdot
B(x)\right)\right\rangle_\alpha\,.\ee
Combining these two equations we find
\be ({\cal D}+\gamma_G)\langle AB\cdots\rangle_\alpha=0\,, \ee
where
\be {\cal D}={\cal
D}_0+2\alpha\gamma_V\frac{\partial}{\partial\alpha}\,. \ee

Next we introduce the gauge field propagator
\be \langle
A^a_\lambda(x),A^b_\mu(y)\rangle=\frac{-i}{(2\pi)^4}\delta_{ab}\int
d^4 k\, e^{ik(x-y)}D_{\lambda\mu}(k)\,, \ee
which is the vacuum expectation value of the time-ordered product of
two color gauge field operators, and
\bea D_{\lambda\mu}(k)&=&\left(\delta_{\lambda\mu}-\frac{k_\lambda
k_\mu}{k^2-i\epsilon}\right)D(k^2,\alpha)+\alpha\frac{k_\lambda
k_\mu}{(k^2-i\epsilon)^2}\,,\label{3.11}\\
D(k^2,\alpha)&=&\int dm^2
\frac{\rho(m^2)}{k^2+m^2-i\epsilon}\,.\label{3.12} \eea
Then introduce
\be R(k^2,\alpha)=k^2\,D(k^2,\alpha)\,.\ee
Since the gauge field has been renormalized in the Landau gauge, we
have
\be R(\mu^2,0)=1.\ee
In other gauges we have to employ the renormalization factor
$Z_3(\alpha)$ instead of $Z_3(0)$, and this amounts to a further
renormalization for $R$. The properly renormalized $R$ function and
the gauge parameter are denoted by $\bar R$ and $\bar\alpha$,
respectively, and are given, in a consistent manner with
(\ref{3.1}), by
\bea R(k^2,\alpha)&=&R(\mu^2,\alpha)\bar R(k^2,\bar\alpha)\,,\\
\alpha&=& R(\mu^2,\alpha)\bar\alpha\,.\eea

The function $R(k^2,\alpha)$ renormalized in the Landau gauge
satisfies
\be({\cal D}+2\gamma_V)R(k^2,\alpha)=0\,.\ee
Then the function $\bar R(k^2,\bar\alpha)$ normalized by
\be \bar R(\mu^2,\bar\alpha)=1\ee
satisfies
\be\label{3.19}({\cal D}+2\bar\gamma_V)\bar
R(k^2,\bar\alpha)=0\,,\ee
where the new anomalous dimension $\bar\gamma_V$ is given by
\be\bar\gamma_V=\gamma_V+\frac{1}{2}{\cal D}\ln
R(\mu^2,\alpha)\,.\ee

Next we switch the set of parameters from $(\mu,g,\alpha)$ to
$(\mu,g,\bar\alpha)$. The anomalous dimension $\bar \gamma_V$ will
be expressed as a function of $g$ and $\bar\alpha$, and it will be
denoted by $\bar \gamma_V(g,\bar\alpha)$ from now on. Then,
\bea {\cal D}&=&({\cal D}\mu)\frac{\partial}{\partial\mu}+({\cal
D}g)\frac{\partial}{\partial g}+({\cal
D}\bar\alpha)\frac{\partial}{\partial\bar\alpha}\cr
&=&\mu\frac{\partial}{\partial\mu}+\beta(g)\frac{\partial}{\partial
g}-2\bar\alpha\,\bar
\gamma_V(g,\bar\alpha)\frac{\partial}{\partial\bar\alpha}\,. \eea
In this way we have established a renormalization prescription
leading to an \a-independent beta-function. For the quark field we
can derive its anomalous dimension $\bar\gamma_\psi$ in a manner
similar to the above derivation of $\bar\gamma_V$. Since the
beta-function is independent of \a, the concept of asymptotic
freedom is gauge-independent. From now on we shall skip the bars
introduced above.

\section{Asymptotic Limits of Running Parameters}
\setcounter{equation}{0}

The RG equations for QCD have been studied in detail and we shall
reinstate their essence in what follows \cite{KN96,NT}.

\vskip 1cm {\it Renormalization Constants}

An element of RG may be expressed as
\be R(\rho)=\exp(\rho{\cal D})\,,\ee
where $\rho$ denotes the parameter of RG and the composition law of
this group is given by
\be R(\rho)R(\rho')=R(\rho+\rho')\,.\ee
Let $Q$ be a function of $g,\ \alpha$ and $\mu$, and we define the
running $Q$ to be
\be\label{4.3} \bar Q(\rho)=\exp(\rho{\cal D})Q(g,\alpha,\mu)=Q(\bar
g(\rho),\bar\alpha(\rho),\bar\mu(\rho))\,,\ee
with the initial condition
\be\bar Q(0)=Q\,.\ee

Let the anomalous dimension of \Gf\ $G(p_i;g,\alpha,\mu)$ be
$\gamma(g,\alpha)$, then we have
\be[{\cal D}+\gamma(g,\alpha)]G(p_i;g,\alpha,\mu)=0\,.\ee
Its running version is defined by
\be\bar G(\rho)=\exp(\rho{\cal D})\cdot G(p_i;g,\alpha,\mu)\,,\ee
and it satisfies
\be \frac{\partial}{\partial\rho}\bar G(\rho)=-\bar\gamma(\rho)\bar
G(\rho)\,.\ee
The formal solution is given by
\be\label{4.8} G(p_i;g,\alpha,\mu)=\exp\left[\int_0^\rho
d\rho'\bar\gamma(\rho')\right]\cdot G(p_i;\bar
g(\rho),\bar\alpha(\rho),\bar\mu(\rho))\,.\ee

Then we assume, in the presence of a cut-off $\Lambda$, that the
running coupling constant $\bar g(\rho)$ tends to the unrenormalized
or the bare one $g_0$ in the limit $\rho \to \infty$, namely,
\be\label{4.9}\lim_{\rho\to\infty}\bar g(\rho)=g^{(0)}\,,\ee
and similarly
\bea\lim_{\rho\to\infty}\bar \alpha(\rho)&=&\alpha^{(0)}\,,\label{4.10}\\
\lim_{\rho\to\infty}\bar
\mu(\rho)&=&\mu\lim_{\rho\to\infty}e^\rho=\infty\,.\label{4.11}\eea
In this limit (\ref{4.8}) reduces to
\be\label{4.12} G(p_i;g,\alpha,\mu)=\exp\left[\int_0^\infty
d\rho'\bar\gamma(\rho')\right]\cdot G^{(0)}(p_i;
g^{(0)},\alpha^{(0)},\infty)\,,\ee
where $G^{(0)}$ denotes the unrenormalized version of \Gf\ $G$. Then
the renormalization constant of $G$ denoted by $Z$ is given by
\be\label{4.13} Z=\exp\left[-\int_0^\infty
d\rho'\bar\gamma(\rho')\right]\,.\ee

The solution of (\ref{3.19}) reads as
\be\label{4.14} R(k^2;g,\alpha,\mu)=\exp\left[2\int_0^\rho
d\rho'\bar\gamma_V(\rho')\right]\cdot R(k^2;\bar
g(\rho),\bar\alpha(\rho),\bar\mu(\rho))\,.\ee
Now apply the Lehmann representation (\ref{3.12}) to the l.h.s. and
take the limit $\rho\to \infty$ after putting $k^2=\bar\mu^2(\rho)$,
then with the help of (\ref{4.13}) we obtain
\be\label{4.15} Z_3^{-1}=\int dm^2\rho(m^2)=\exp\left[2\int_0^\infty
d\rho'\bar\gamma_V(\rho')\right]\,.\ee
In the cut-off theory we first take the limit  $\rho\to\infty$ and
then $\Lambda\to\infty$, but in what follows we invert the order of
limiting procedures by taking the limit $\Lambda\to\infty$ first.
Thus some of the initial conditions introduced in the cut-off theory
are not necessarily satisfied. As an example we shall see later that
the limiting values (\ref{4.9}) and (\ref{4.10}) cannot be arbitrary
despite our expectation that the unrenormalized $g^{(0)}$ and
$\alpha^{(0)}$ should be chosen arbitrarily.

\vskip 1cm{\it Asymptotic Limits of Running Parameters}

Next we shall study the RG equations for the running parameters
$\bar g(\rho),\ \bar\alpha(\rho)$ and $\bar\mu(\rho)$ that follow
from (\ref{4.3}),
\bea \frac{d}{d\rho}\bar g(\rho)&=&\bar \beta(\rho)\,,\label{4.16}\\
\frac{d}{d\rho}\bar\alpha(\rho)&=&-2\bar
\alpha(\rho)\bar\gamma_V(\rho)\,,\cr
\frac{d}{d\rho}\bar \mu(\rho)&=&\bar \mu(\rho)\ \ \mbox{or}\ \ \bar
\mu(\rho)=\mu\, e^{\rho}\,.\nonumber \eea
First we shall define their asymptotic limits by
\be \bar g(\infty)=g_\infty,\ \ \bar\alpha(\infty)=\alpha_\infty,\ \
\bar\mu(\infty)=\infty\,.\ee
In the absence of a cut-off they do not necessarily reduce to their
unrenormalized counterparts in (\ref{4.9}) and (\ref{4.10}).

In perturbation theory $\beta(g)$ and $\gamma_V(g,\alpha)$ are given
in the form of power series,
\bea \beta(g)&=&g^3(\beta_0+\beta_1\,g^2+\cdots)\,,\label{4.18}\\
\gamma_V(g,\alpha)&=&g^2(\gamma_0(\alpha)+\gamma_1(\alpha)\,g^2+\cdots)\,,\label{4.19}
 \eea
where
\bea \gamma_0(\alpha)&=&\gamma_{00}+\gamma_{01}\alpha\,,\cr
\gamma_1(\alpha)&=&\gamma_{10}+\gamma_{11}\alpha+\gamma_{12}\alpha^2\,,\
\ \cdots\,,\label{4.20}
 \eea
The lowest order coefficients are given by
\bea\label{4.21}
\beta_0&=&-\frac{1}{32\pi^2}\left(22-\frac{4}{3}N_f\right)\,,\cr
\gamma_{00}&=&-\frac{1}{32\pi^2}\left(13-\frac{4}{3}N_f\right)\,,\\
\gamma_{01}&=&\frac{3}{32\pi^2}>0\,.\nonumber \eea
When $\beta_0$ is negative, namely, when $N_f\leq16$, asymptotic
freedom is realized and we shall assume it in what follows.

Asymptotic freedom \cite{Gross, Politzer} is characterized by
\be g_\infty=0\,,\ee
and for large values of $\rho$ we obtain approximately
\be\label{4.23}\bar g^2(\rho)=\frac{1}{b\rho}\ \ \
(b=-2\beta_0>0)\,.\ee
By integrating the second equation in (\ref{4.16}) we find a sum
rule \cite{KN96, NT}:
\be\ln\frac{\alpha_\infty}{\alpha}=-2\int_0^\infty
d\rho\bar\gamma_V(\rho)\,,\label{4.24}\ee
or
\be Z_3^{-1}=\exp\left[2\int_0^\infty
d\rho\bar\gamma_V(\rho)\right]=\frac{\alpha}{\alpha_\infty}\,.\label{4.25}\ee
The essence of our argument is based on this sum rule.

Our main problem is the determination of the asymptotic limit
$\alpha_\infty$, but this problem has been discussed in detail
before, so we shall only quote the results in what follows. We shall
come back to it, however, in the next section for a different
purpose.

For $\alpha=0$ we have $\bar\alpha(\rho)=0$ and hence
$\alpha_\infty=0$. It is clear that $\bar\alpha(\rho)$ and $\alpha$
are always of the same signature. We find three possibilities for
$\alpha_\infty$:
\be\alpha_\infty=-\infty,\ 0,\ \alpha_0\,,\label{4.26}\ee
where $\alpha_0$ is defined by
\be\label{4.27}\gamma_0(\alpha_0)=\gamma_{00}+\gamma_{01}\alpha_0=0\
\ \mbox{or}\ \ \ \alpha_0=-\gamma_{00}/\gamma_{01}\,.\ee

For $\alpha_0>0$ or $N_f<10$, we find
\begin{equation}\label{4.28}
\alpha_\infty =\left \{\begin{tabular}{ll}
$\alpha_0\,,$ &  $\alpha>0$ \\
0\,, & $\alpha=0$\\
$-\infty\,,$ &  $\alpha<0\,,$\\
\end{tabular}
\right.
\end{equation}
and in this case the sum rule can be expressed as
\be\int
dm^2\,\rho(m^2)=\frac{\alpha}{\alpha_\infty}\theta(\alpha)\label{4.29}\ee
where
\begin{equation}\label{4.30}
\theta(\alpha) =\left \{\begin{tabular}{ll}
$1\,,$ &  $\alpha\geq0$ \\
0\,, & $\alpha<0$\,.\\
\end{tabular}
\right.
\end{equation}
The r.h.s. of (\ref{4.29}) is a continuous function of $\alpha$, but
its derivative with respect to $\alpha$ develops a discontinuity at
$\alpha=0$. Thus, the set of real $\alpha$, denoted by $D$, is
divided into three connected subsets:
\be\label{4.31}D=D(-\infty)\bigcup D(0)\bigcup D(\alpha_0)\,,\ee
where we confine ourselves to positive $\alpha_0$ or $N_f<10$
\bea D=\{\alpha|\ \ \alpha\in{R}\}\,,\cr
D(-\infty)=\{\alpha|\ \ \alpha<0\}\,,\cr
D(0)=\{\alpha|\ \ \alpha=0\}\,,\cr
D(\alpha_0)=\{\alpha|\ \ \alpha>0\}\,.
 \eea
Thus, for gauge-dependent \Gfs\ such as the gluon propagator we find
three equivalence classes of gauges.

In one of the papers by the present authors \cite{CN06} the residue
of the massless pole in the two-point function
\be\label{4.33}\langle A_\lambda^a(x),F_{\mu\nu}^b(y)\rangle\ee
has been studied, and it has been concluded that the residue
vanishes in two classes $D(-\infty)$ and $D(0)$, suggesting that the
gluon in these gauges would likely be massive, whereas it is
non-vanishing in the class $D(\alpha_0)$ implying zero mass for the
gluon. This means that the gluon mass would be gauge-dependent. The
interpretation of this result is delicate. In that paper we tacitly
assumed that the gluon mass is a physical quantity and that the
above result indicates the gauge-dependence of QCD, or more
precisely the class-dependence of the gluon mass, namely, it is zero
in the equivalence class of gauges $D(\alpha_0)$, but it is non-zero
in the classes $D(-\infty)\bigcup D(0)$. The gluon mass denoted by
$M$ must be RG invariant
\be\label{4.35}{\cal D} M=0\ee
within a class.

We found that this interpretation is misleading since the physically
observable quantities must be gauge-independent as illustrated by
(\ref{2.29}).

The $S$-matrix elements for hadronic processes are gauge-independent
as is clear from (\ref{2.29}), and the condition for color
confinement \cite{KN96,NT,KN94,KN95,CN01}
\be\label{4.36}Z_3^{-1}=0\ee
is satisfied in the classes $D(-\infty)\bigcup D(0)$. Therefore, in
these classes color confinement is realized and the unitarity
condition of the $S$-matrix between two hadronic states $|a\rangle$
and $|b\rangle$ reads as
\be\label{4.37}\sum_n\langle b|S^\dagger|n\rangle\langle
n|S|a\rangle=\langle b|a\rangle\,,\ee
where the sum over the intermediate states is saturated by hadronic
states alone without introduction of the confined quarks and gluons.
Since the $S$-matrix is gauge-independent this statement is also
valid for the class $D(\alpha_0)$. Thus, confinement is a
gauge-independent or class-independent concept. This conclusion does
not contradict the gauge-dependence of the gluon mass since the mass
of the confined gluons is never observed and turns out to be an
unphysical quantity. A similar observation has been made by
Fujikawa, Lee and Sanda \cite{Fujikawa}, that particles with
gauge-dependent masses are unphysical and not subject to observation
in connection with the $R_\xi$ gauge.

\section{Introduction of the Complex Gauge Parameter}
\setcounter{equation}{0}

From (\ref{4.19}) and (\ref{4.20}) we can readily deduce  that
$\gamma_V$ can be expanded into a double power series in $g^2$ and
$\alpha g^2$. For sufficiently large $\rho$, we can assume that
$\bar g^2(\rho)\ll1$ and if $\bar\alpha(\rho)$ is bounded below a
certain constant, we may use perturbation theory in powers of $g^2$.
In the lowest order we have
\be\label{5.1}\frac{d\bar g}{d\rho}=\beta(\bar g)\approx\beta_0\bar
g^3=-\frac{b}{2}\bar g^3\,,\ee
and its solution is given by
\be\label{5.2}\bar g^2(\rho)=\frac{g^2}{1+b\,g^2\,\rho}\,.\ee
When $\bar\alpha$ is bounded we employ the following approximate
equation:
\be\label{5.3}\frac{d\bar\alpha}{d\rho}=-2\bar\alpha\bar\gamma_V\approx
f(\rho)\bar\alpha(\alpha_0-\bar\alpha)\,,\ee
where
\be\label{5.4}f(\rho)=2\,\gamma_{01}\,\bar g^2(\rho)>0\,.\ee
When the integral (\ref{4.24}) is convergent, $\alpha_\infty$ must
be finite. The expansion of $\gamma_V$ in powers of $g^2$ starts
from $g^2$ and for large values of $\rho$ the behavior of $\bar
g^2(\rho)$ is given by (\ref{4.23}) so that the integral of the
first term in the expansion of $\bar\gamma_V$ diverges like
\be\label{5.5}\int_0^\infty d\rho\,\bar
g^2(\rho)\sim\frac{1}{b}\ln\infty\,.\ee

In order for the integral of the power series to converge term by
term in (\ref{4.24}), therefore, the condition
\be\label{5.6}\gamma_0(\alpha_\infty)=0\ee
must be satisfied. With reference to (\ref{4.27}) we find
\be\label{5.7}\alpha_\infty=\alpha_0\ee

When the integral (\ref{4.24}) is divergent, $\ln
(\alpha/\alpha_\infty)$ should diverge so that we find
\be\label{5.8}\alpha_\infty=0\ \ \mbox{or}\ \ \pm\infty\,.\ee
Now, when $\bar\alpha$ goes off $\alpha_\infty$ starting from its
neighborhood for increasing $\rho$, $\alpha_\infty$ is called a
repulsive asymptotic limit. Otherwise, when $\bar\alpha$ approaches
$\alpha_\infty$ again starting from its neighborhood, it is called
an attractive asymptotic limit.

\vskip 1 cm{\it Case of positive $\alpha_0$} ($N_f<10$)

Integration of (\ref{5.3}) in the neighborhood of $\alpha_0$ yields
\be\label{5.9}\frac{\alpha_0-\bar\alpha(\rho)}{\alpha_0-\alpha}\cong\exp\left[-\alpha_0\int_0^\rho
d\rho'\,f(\rho')\right]\to 0\,,\ \ \ \mbox{for}\ \ \
\rho\to\infty\,.\ee
Thus $\alpha_0$ is found to be attractive. In the neighborhood of 0,
on the other hand, we find
\be\label{5.10}\frac{\bar\alpha(\rho)}{\alpha}\cong\exp\left[\alpha_0\int_0^\rho
d\rho'\,f(\rho')\right]\to \infty\,,\ \ \ \mbox{for}\ \ \
\rho\to\infty\,.\ee
Thus $\alpha_\infty=0$ is found to be repulsive.

The flow of $\alpha(\rho)$ on the real $\alpha$ axis for increasing
$\rho$ is given in Fig. 1.

\vskip0.3cm
\begin{figure}[h!]
\begin{center}
\includegraphics[width=10cm] {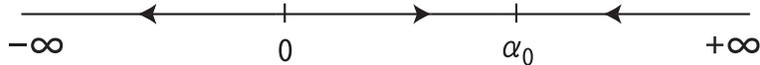}
\end{center}
\caption{Flow of $\bar\alpha(\rho)$ along the real $\alpha$ axis for
increasing $\rho$ ($\alpha_0>0$).} \label{fig1}
\end{figure}

From Fig. 1 we conclude that $-\infty$ is also attractive.

\vskip 1cm{\it Case of negative $\alpha_0$} ($10\leq N_f\leq16$)

In this case 0 and $-\infty$ are attractive and $\alpha_0$ is
repulsive.

\begin{figure}[h!]
\begin{center}
\includegraphics[width=10cm] {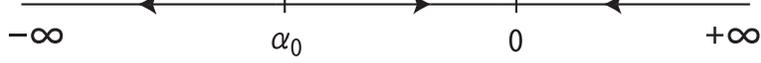}
\end{center}
\caption{Flow of $\bar\alpha(\rho)$ along the real $\alpha$ axis for
increasing $\rho$ ($\alpha_0<0$).} \label{fig2}
\end{figure}

In what follows we shall confine ourselves to the case of positive
$\alpha_0$ and shall study what would take place when $\alpha$ turns
out to be complex while keeping $g^2$ real. For this purpose we
assume that $g^2$ is already sufficiently small and that
$|\bar\alpha(\rho)|$ is bounded to guarantee the following treatment
of the RG equations.

The solution of (\ref{5.3}) as combined with (\ref{5.2}) is given by
\be\label{5.11}\bar\alpha(\rho)=\alpha_0\left[\left(\frac{\alpha_0}{\alpha}-1\right)(1+b\,g^2\,\rho)^{-n}+1\right]^{-1}\,,\ee
where
\be\label{5.12}n=\gamma_{00}/\beta=\left(13-\frac{4}{3}N_f\right)/\left(22-\frac{4}{3}N_f\right)\,.\ee
For $\alpha_0>0$ we have $n>0$.

\vskip 1cm

(1) $\alpha>0$

In this case we can easily check for  $\rho>0$ that
\be\label{5.13}\left(\frac{\alpha_0}{\alpha}-1\right)(1+b\,g^2\,\rho)^{-n}+1>0\,,\ee
and, as has been expected, we have
\be\label{5.14}\lim_{\rho\to\infty}\bar\alpha(\rho)=\alpha_0\,.\ee

\vskip 1cm

(2) $\alpha=0$

In this case, $\alpha=0$ is a fixed point, so that we have
$\bar\alpha(\rho)=0$, and consequently
\be\label{5.15}\lim_{\rho\to\infty}\bar\alpha(\rho)=0\,.\ee

\vskip 1cm

(3) $\alpha<0$

The asymptotic limit of $\bar\alpha(\rho)$, when we start from a
negative $\alpha$, should be equal to either 0 or $-\infty$. Since
$\alpha_\infty=0$ is repulsive, $\alpha_\infty=-\infty$ is the only
choice. In this case we cannot use the series expansion in powers of
$\alpha$ since $|\bar\alpha|$ increases indefinitely for increasing
$\rho$.

\vskip 1cm

(4) complex $\alpha$

Give an imaginary part $i\epsilon$ to be added to $\alpha$
\be\label{5.16}\alpha\to\alpha+i\epsilon\,,\ee
then
\be\label{5.17}\left(\frac{\alpha_0}{\alpha+i\epsilon}-1\right)(1+b\,g^2\,\rho)^{-n}+1\neq0\,,\ee
so that the formula (\ref{5.11}) gives
\be\label{5.18}\lim_{\rho\to\infty}\bar\alpha(\rho)=\alpha_0\,,\ \ \
(\epsilon\neq0)\ee
for an arbitrary choice of the real part $\alpha$, provided that the
imaginary part is non-zero. As mentioned in (1) this result is also
valid even for $\epsilon=0$ for $\alpha>0$. The only exceptions are
the cases $\alpha=0$ and $\alpha<0$ as mentioned in (2) and (3).

In Section 4 we have divided the set of real $\alpha$ into three
subsets or three classes on the real axis, but we can extend this
division to the set of complex $\alpha$ as
\be\label{5.19}D^{(2)}=D^{(1)}(-\infty)\bigcup D^{(0)}(0)\bigcup
D^{(2)}(\alpha_0)\,,\ee
where
\bea\label{5.20} D^{(2)}&=&\{\alpha|\ \ \alpha\in{C}\}\,,\cr
D^{(1)}(-\infty)&=&\{\alpha|\ \ \alpha<0\}\,,\cr
D^{(0)}(0)&=&\{\alpha|\ \ \alpha=0\}\,,\cr
D^{(2)}(\alpha_0)&=&\{\alpha|\ \ \alpha\in C\,,\ \ \alpha\leq0\ \
\mbox{excluded}\}\,.
 \eea
The superscripts $(2),\ (1)$ and $(0)$ denote the dimensionality of
these sets or classes, respectively.

Next we shall generalize Fig. 1 to the complex plane. The lines of
RG flow show their behavior qualitatively or topologically, but not
quantitatively.

\begin{figure}[h!]
\begin{center}
\includegraphics[width=10cm] {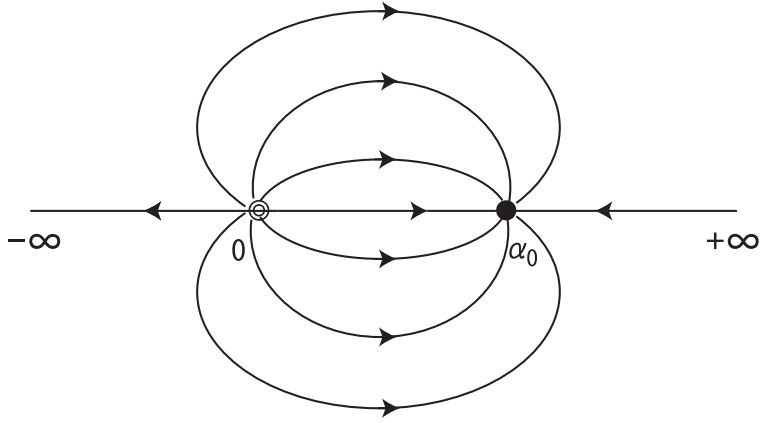}
\end{center}
\caption{Lines of RG flow in the complex $\alpha$ plane for
increasing $\rho$ ($\alpha_0>0$).} \label{fig3}
\end{figure}

It is interesting to recognize that the lines of RG flow resemble
the lines of force generated by a dipole. All lines of flow tend to
$\alpha_0$, except for the one along the negative real axis that
cannot go up nor down in the complex plane.

In (\ref{4.29}) we found that $\alpha=0$ is a discontinuity for the
integral of the spectral function of the gluon propagator. We can
extend this observation to the complex $\alpha$ plane, and for this
purpose we shall write down the sum rule for negative $\alpha$ and
also for $\alpha+i\epsilon$:
\bea\int dm^2\,\rho(m^2,g^2,\alpha,\mu)&=&0\,,\ \ \
(\alpha<0)\label{5.21}\\
\lim_{\epsilon\to 0}\int
dm^2\,\rho(m^2,g^2,\alpha+i\epsilon,\mu)&=&\frac{\alpha}{\alpha_0}\,,\
\ \ (\alpha<0)\,. \eea
This shows that there is a discontinuity in the integral of the
spectral function when $\alpha$ approaches the negative real axis.

\vskip 1cm {\it QED}

The situation is completely different in QED. First of all, a
gauge-invariant two-point function
\be\label{5.23}\langle F_{\lambda\sigma}(x),F_{\mu\nu}(y)\rangle\ee
can be expressed in terms of the transverse part of the photon
propagator so that the photon mass that appears as the pole of this
expression is physical. QED is characterized by Ward's identity
\be\label{5.24}Z_1=Z_2\,,\ee
which is expressed in terms of anomalous dimensions by
\be\label{5.25}\beta(e)=e\gamma_V(e)\,,\ee
from which we can derive
\be\label{5.26}\frac{d}{d\rho}[\bar
e^2(\rho)\,\bar\alpha(\rho)]=0\,.\ee
As mentioned already, the spectral function $\rho(m^2)$ does not
depend on $\alpha$, nevertheless, we have the sum rule:
\be\label{5.27}Z_3^{-1}=\int
dm^2\,\rho(m^2)=\frac{\alpha}{\alpha_\infty}>1\,.\ee
The only consistent choice of $\alpha_\infty$ that makes $Z_3$
independent of $\alpha$ is $\alpha_\infty=0$, and
$e^2_\infty=\infty$ follows from (\ref{5.26}).

In this case, $\alpha_\infty$ is the only asymptotic limit that is
attractive, and there is only one equivalence class of gauges in
QED. Furthermore, $e_\infty=\infty$ signifies that $\gamma_V(e)$ is
positive definite, and the expression
\be\label{5.28}\frac{\bar\alpha(\rho)}{\alpha}=\exp\left[-2\int_0^\rho
d\rho'\,\bar\gamma_V(\rho')\right]\ee
is real and decreases with increasing $\rho$. Then, the lines of RG
flow in QED are shown in Fig. 4.

\begin{figure}[h!]
\begin{center}
\includegraphics[width=10cm] {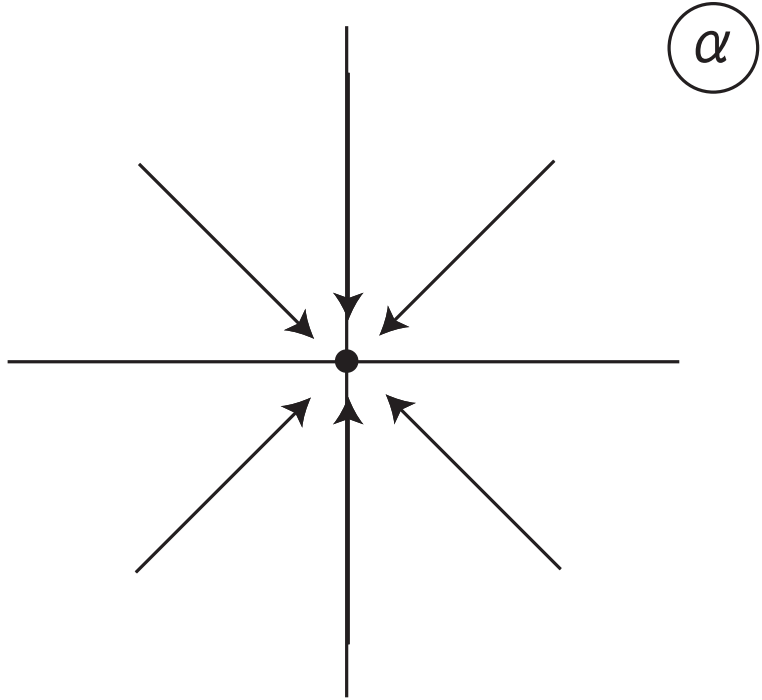}
\end{center}
\caption{Lines of RG flow in the complex $\alpha$ plane for
increasing $\rho$ (QED).} \label{fig4}
\end{figure}

Here the lines of RG flow resemble the lines of force generated by a
monopole.

In QED we have only one equivalence class of gauges and the massless
photon is physical and observable. In this case we have
\be\label{5.29}Z_3^{-1}=\infty\ee
and charge confinement is not realized.

\section{Conclusions}

We may summarize what we have done in this paper in connection with
the results obtained in a series of papers on this subject
\cite{KN96,NT,KN94,KN95}.

\vskip 0.5cm

1) Color $SU(3)$ symmetry

For color confinement we need an exact non-Abelian gauge symmetry.
If this symmetry is broken, a color singlet state is forced to mix
with colored states and consequently color cannot be confined.
Mathematically, the original form of the condition for color
confinement was given by the absence of the massless spin zero
component in the current $\delta\bar\delta A_\lambda^a$, but
breaking of color symmetry would induce the unwanted massless spin
zero component in that current in the form of the Nambu-Goldstone
boson. In Abelian gauge theories such as QED the massless spin zero
component is always present, thereby preventing the charges from
being confined.

\vskip 0.5cm

2) Condition for Color Confinement

With the help of renormalization group we can show that a sufficient
condition for color confinement can be cast in the form of Eq.
(\ref{4.36}), which formed the basis of our discussion in the
present paper.

\vskip 0.5cm

3) Evaluation of $Z_3$

Furthermore, with the help of renormalization group we can derive
the identity (\ref{4.25}) and can show with the help of asymptotic
freedom that for negative choices of $\alpha$, eventually below a
certain negative constant, $\alpha_\infty$ turn out to be $-\infty$,
satisfying the condition (\ref{4.36}) for confinement.

\vskip 0.5cm

4) $S$-matrix in Confined QCD

By using gauge-independent \Gfs\ we can express the $S$-matrix
elements for hadronic reactions by applying the reduction formula
\cite{LSZ}. The unitarity condition for the $S$-matrix between two
hadronic states are expressed by Eq. (\ref{4.37}), and confinement
is characterized by the statement that the sum over the intermediate
states is saturated by keeping only hadronic states, excluding
colored particles such as quarks and gluons.

If confinement is realized in one gauge in the above sense, it is
clearly realized also in all other gauges, since the $S$-matrix is
gauge-independent. In this sense the concept of gauge-independence
is a key factor in understanding confinement. The importance of
gauge-independence or invariance has been stressed by many authors
\cite{Fujikawa,tHooft,Feynman}.

\vskip 1cm {\bf{Acknowledgements}}

One of the authors (K.N.) is grateful to D.V. Shirkov for useful
discussion on the subject and acknowledges the support by a
Grant-in-Aid for Scientific Research from MEXT of Japan. Last but
not least, we owe M. Chaichian his constant encouragement in
pursuing this work.

\end{document}